\documentclass[aps,prd,superscriptaddress,floatfix,showpacs]{revtex4}

\usepackage{verbatim}

\usepackage{amsmath}
\usepackage{amsfonts}
\usepackage{amssymb}
\usepackage{graphicx}
\usepackage{tikz-feynman}
\usepackage{slashed}
\usepackage{bm}
\usepackage{color}
\usepackage{epsf}

\usepackage{psfrag}

\def\bea{\begin{eqnarray}}
\def\eea{\end{eqnarray}}
\def\beas{\begin{eqnarray*}}
\def\eeas{\end{eqnarray*}}
\def\beqas{\begin{eqnarray*}}
\def\eqas{\end{eqnarray*}}
\def\beq{\begin{equation}}
\def\eeq{\end{equation}}
\def\beqd{\begin{displaymath}}
\def\eeqd{\end{displaymath}}
\def\eqd{\end{displaymath}}

\def\slashchar#1{\setbox0=\hbox{$#1$}
   \dimen0=\wd0
   \setbox1=\hbox{/} \dimen1=\wd1
   \ifdim\dimen0>\dimen1
      \rlap{\hbox to \dimen0{\hfil/\hfil}}
      #1
   \else\begin{eqnarray}
      \rlap{\hbox to \dimen1{\hfil$#1$\hfil}}
      /
   \fi}

\begin{document}

\title
{Collinear factorization of diphoton photoproduction at next to leading order}
\author{O.~Grocholski}
\affiliation{ National Centre for Nuclear Research (NCBJ), 02-093 Warsaw, Poland}
\affiliation{Institute of Theoretical Physics, Faculty of Physics, University of Warsaw, Pasteura 5, 02-093
Warsaw, Poland}
\author{ B.~Pire}
\affiliation{ Centre de Physique Th\'eorique, CNRS, École Polytechnique, I.P. Paris, 91128 Palaiseau, France  }
\author{ P.~Sznajder}
\affiliation{ National Centre for Nuclear Research (NCBJ), 02-093 Warsaw, Poland}
\author{ L.~Szymanowski}
\affiliation{ National Centre for Nuclear Research (NCBJ), 02-093 Warsaw, Poland}
\author{ J.~Wagner}
\affiliation{  National Centre for Nuclear Research (NCBJ), 02-093 Warsaw, Poland}
\date{\today}
\begin{abstract}
 We calculate in the framework of collinear factorization the amplitude for the photoproduction of a near forward large mass diphoton at leading twist and next to leading order (NLO) in $\alpha_s$. We demonstrate the validity of factorization at this order, which was never achieved for such a reaction where the coefficient function describes a $2 \to 3$ hard process. While the Born order amplitude was purely imaginary and only probed the $x=\pm \xi$ cross-over line of generalized parton distributions (GPD) domain, the NLO result contains both a real and an imaginary part and probes the whole domain of definition of quark GPDs. The phenomenology of our results for medium (JLab) and higher energy (EIC) experiments will be developed in a  future study.
\end{abstract}
\maketitle


\section{Introduction}
In our recent works \cite{Pedrak:2017cpp, Pedrak:2020mfm} we studied at leading order (LO) in the QCD coupling constant $\alpha_s$ the photoproduction of a large mass diphoton on a    nucleon target
\begin{equation}
\gamma(q,\epsilon) + N(p_1,s_1) \rightarrow \gamma(q_1,\epsilon_1) +  \gamma(q_2,\epsilon_2)+ N'(p_2,s_2)\,,
\label{process}
\end{equation}
 in the kinematical domain suitable to a factorization of generalized parton distributions (GPDs) and a hard amplitude. 
 This domain is defined by large 
diphoton invariant mass and center of mass energy and by a small momentum transfer squared $t =(p_2-p_1)^2$ between the initial and  final nucleons. 
This reaction, which superficially looks like timelike Compton scattering (TCS) \cite{Mueller:1998fv, Berger:2001xd, Pire:2011st,Moutarde:2013qs, CLAS:2021lky} with the dilepton replaced by a diphoton, has a quite different structure since the hard amplitude 
describes a $2 \to 3$ process, contrarily to more familiar reactions 
 -- deeply virtual Compton scattering (DVCS), deeply virtual meson production (DVMP) or TCS -- which involve 
 a $2 \to 2$ process. While other  $2 \to 3$ processes have already been studied at leading order \cite{Ivanov:2002jj,Kumano:2009he,Beiyad:2010cxa,Boussarie:2016qop,Duplancic:2018bum}, assuming QCD collinear factorization at leading twist, there is yet no proof of this very basic property, even at the next to leading  order (NLO) in $\alpha_s$.
  
 It is thus deemed appropriate to calculate at NLO in  $\alpha_s$ the amplitude of the simplest $2 \to 3$ process, namely the photoproduction of a diphoton (\ref{process}). This is the aim of this paper where we  demonstrate the factorization at NLO of the amplitude, showing in particular that the divergent parts of the NLO amplitude can be absorbed in the GPDs. We use dimensional regularization as the main tool of our proof. The phenomenological 
consequences of our NLO description of process (\ref{process}) will be analyzed in a forthcoming publication.
 
In Sect. II we present the kinematics of our process, then we briefly recall in Sect. III  the structure of a factorization proof at NLO. Section IV discusses in some details the calculation of the vector GPD contribution 
for our process. In Section V we discuss the corresponding results for the axial amplitude. Section VI contains some remarks and conclusions. Some of the results of this paper have been already 
presented in~\cite{Grocholski:2021dgi,Grocholski:2021gta}. 
\section{kinematics}
\label{sec_kinematics}
We use the following light-cone four-vectors:
\begin{equation}
p^\mu = \frac{\sqrt{s}}{2}(1,0,0,1), \quad \quad n^\mu =  \frac{\sqrt{s}}{2}(1,0,0,-1),
\end{equation}
the parameter $s= 2 p\cdot n$ will be described in more detail later.
We parametrize 
the momenta in the Sudakov basis:
\begin{equation}
\begin{aligned}
&p_1^\mu = (1+\xi)p^\mu + \frac{M^2}{s(1+\xi)}n^\mu, \qquad p_2^\mu = (1-\xi)p^\mu + \frac{M^2 + \vec{\Delta}_\perp^2}{s(1-\xi)}n^\mu + \Delta_\perp^\mu, \qquad q^\mu = n^\mu,
\\
&q_1^\mu = \alpha n^\mu + \frac{(\vec{p}_\perp-\frac{1}{2}\vec{\Delta}_\perp)^2}{\alpha s} p^\mu + p_\perp^\mu - \frac{1}{2}\Delta_\perp^\mu,
\\
&q_2^\mu = \bar{\alpha} n^\mu + \frac{(\vec{p}_\perp+\frac{1}{2}\vec{\Delta}_\perp)^2}{\bar{\alpha} s} p^\mu - p_\perp^\mu - \frac{1}{2}\Delta_\perp^\mu.
\end{aligned}
\end{equation}
Here, $M$ is the mass of the nucleon, $\xi$ is the skewness variable and $\vec{v}\vec{u}$ denotes the Euclidean product of transverse vectors. 
Polarization vectors of the photons are written in the $\epsilon \cdot p =0$ gauge 
as:
\begin{equation}
\begin{aligned}
&\epsilon^\mu(q_1) = \epsilon_\perp^\mu(q_1) - \frac{2\epsilon_\perp\big{(}p_\perp - \frac{1}{2}\Delta_\perp\big{)}}{\alpha s}p^\mu,
\\
&\epsilon^\mu(q_2) = \epsilon_\perp^\mu(q_2) + \frac{2\epsilon_\perp\big{(}p_\perp + \frac{1}{2}\Delta_\perp\big{)}}{\bar{\alpha} s}p^\mu,
\\
&\epsilon^\mu(q) = \epsilon_\perp^\mu(q).
\end{aligned}
\end{equation}
The process can be described in terms of the following $4$ invariants:
\begin{equation}
S_{\gamma N} = (p_1 + q)^2, \quad t = (p_1 - p_2)^2, \quad M_{\gamma\gamma}^2 = (q_1 + q_2)^2, \quad u' = (q_2 - q)^2.
\end{equation}
For further convenience we relate them to other invariants:
\begin{equation}
\begin{aligned}
&S_{\gamma N} = (1+\xi) s + M^2 \: \rightarrow \: s = \frac{S_{\gamma N} - M^2}{1+\xi},
\\
&t' = (q_1 - q)^2 = t - M_{\gamma\gamma}^2 - u'.
\end{aligned}
\end{equation}
At leading twist, the calculation of the hard part uses the
approximation $\Delta_\perp = 0$, justified by the fact that in the considered kinematics one has $|p_\perp^2| \gg |\Delta_\perp^2|$. The simplified kinematical relations  then read:
\begin{equation}
\begin{aligned} \label{eq-kin-simplified}
& \tau = \frac{M_{\gamma\gamma}^2 - t}{S_{\gamma N} - M^2}, \qquad \xi = \frac{\tau}{2-\tau}, \\
&-t' = \frac{\vec{p}_\perp^{\:2}}{\alpha}, \qquad -u' = \frac{\vec{p}_\perp^{\:2}}{\bar{\alpha}}, \qquad \alpha + \bar{\alpha} = 1.
\end{aligned}
\end{equation}

To reduce the number of considered diagrams, we will consider a process with three incoming photons with momenta $(k_1,k_2,k_3)$ and polarisation vectors $(\epsilon_1, \epsilon_2, \epsilon_3)$, 
and then sum the amplitudes over $\big{\{}(k_1,\epsilon_1), (k_2, \epsilon_2) ,(k_3,\epsilon_3)\big{\}}$ corresponding to all permutations of $\big{\{}(-q_1,\epsilon^* (q_1)), (-q_2, \epsilon^* (q_2) ) ,(q,\epsilon(q) )\big{\}}$.
We introduce the parameters $\beta_i$, $\kappa_i$ defined as: 
\begin{equation}
2pk_i := \beta_i s, \quad 2k_1 k_2 := \kappa_3 s, \quad 2k_2 k_3 := \kappa_1  s, \quad 2k_1 k_3 := \kappa_2  s.
\end{equation}
It is easy to check that $\kappa_i = 2\xi \beta_i.$
Scalar products of corresponding momenta are:
\begin{equation}
-2pq_1 = -\alpha s, \qquad -2pq_2 = -\bar{\alpha} s, \qquad 2pq = s.
\end{equation}
Adding the different diagrams amounts to let the  $\beta_i$ and $\kappa_i$ parameters  
be subject to the following permutations: 
\begin{equation}
\begin{aligned}
&\{\beta_i\} = \: \mathrm{perm.} \: \mathrm{of} \: \Big\{ -\alpha, - \bar{\alpha}, 1 \Big\}\,,
\\
&\{\kappa_i\} = \: \mathrm{perm.} \: \mathrm{of} \: \Big\{ \frac{u'}{s} ,\frac{t'}{s} , \frac{M_{\gamma\gamma}^2}{s}\Big\}\,.
\end{aligned}
\end{equation}

In the calculation of the hard part it is assumed that $p_1-p_2 \approx 2\xi p$, so that $2\xi p = q_1 + q_2 - q$. Hence, for each polarization vector $\epsilon_i$:
\begin{equation}
\epsilon_i (k_1 + k_2 + k_3 ) = 0.
\end{equation}
Moreover, $k_1 \epsilon_1 = k_2 \epsilon_2= k_3 \epsilon_3 =0$. 
In the considered gauge, only the transverse parts of polarizations vectors contribute to their products with momenta. We have:
\begin{equation}
\begin{aligned}
&q_1 \epsilon(q) = -q_2\epsilon(q) = -\vec{p}_t \vec{\epsilon}_t(q),
\\
&q \epsilon(q_1) = q_2 \epsilon(q_1) = \frac{1}{\alpha} \vec{p}_t \vec{\epsilon}_t(q_1), \qquad q \epsilon(q_2) = q_1 \epsilon(q_2) =  -\frac{1}{\bar{\alpha}} \vec{p}_t \vec{\epsilon}_t(q_2).
\end{aligned}
\end{equation}

%
%
\section{Factorization}
In the collinear factorization regime, 
the scattering amplitude of our process (\ref{process}) reads:
\begin{eqnarray}
\mathcal{A} &=& \sum_q\int_{-1}^1 dx \left[
 T_B^q(x) F_B^q(x)  
+ \widetilde{T}_B^q(x) \widetilde{F}_B^q(x)
 \right]
 \,,
\label{eq:factorizedamplitude}
\end{eqnarray}
where $F_B^q$ and $\widetilde{F}_B^q$ are bare vector and axial quark GPDs; $T_B^q$ and $\widetilde{T}_B^q$ are bare vector and axial coefficient functions containing infrared divergences at $O(\alpha_s)$, and are written as:
\begin{eqnarray}
T_B^q &=&Tr\left[\mathcal{M}\frac{\not\!{p}}{4}\right]=
C_0^q + 
\left(
\frac{M^2_{\gamma\gamma}e^\gamma}{4\pi\mu_R^2}
\right)^{-\epsilon/2}
\left(
-\frac{2}{\epsilon}~C_{coll}^{q} + C_1^{q}
\right) \label{eq-def-Ccoll} \,,\\
\tilde{T}_B^q &=&Tr\left[\mathcal{M}\gamma^5\frac{\not\!{p}}{4}\right]=
\tilde{C}_0^q + 
\left(
\frac{M^2_{\gamma\gamma}e^\gamma}{4\pi\mu_R^2}
\right)^{-\epsilon/2}
\left(
-\frac{2}{\epsilon}~\widetilde{C}_{coll}^{q} + \widetilde{C}_1^{q}
\right)\,,
\label{eq:coeff}
\end{eqnarray}
where $\mathcal{M}$ is the $q\gamma \to q \gamma\gamma$ hard scattering amplitude with the quark spinors removed.

Only nonsinglet quark GPDs contribute to our process and the $O(\alpha_s)$ relation between the bare and renormalized GPDs in the $\overline{\mathrm{MS}}$ scheme reads:
\begin{eqnarray}
F_B^q(x) &=& F^q(x) 
+ \left(\frac{2}{\epsilon} +\ln\frac{e^\gamma \mu_F^2}{4\pi\mu_R^2} \right) K^{qq}_{NS}(x,x')\otimes F^q(x')\,,
\label{eq:evolution}
\end{eqnarray}
where $\mu_F$ and $\mu_R$ are respectively the factorization and renormalization scales, and $\gamma$ is the Euler constant. The cancellation of divergences - a necessary requirement for a consistent factorization framework - occurs, if:
\begin{eqnarray}
C_{coll}^{q}(x') &=& C_0^q(x) \otimes K_{NS}^{qq}(x,x')~~,~~ \widetilde C_{coll}^{q}(x') = \widetilde C_0^q(x) \otimes K_{NS}^{qq}(x,x'),
\label{eq:div_cancel}
\end{eqnarray}
which leads to:
\begin{eqnarray}
\mathcal{A} &=& \sum_q\int_{-1}^1 dx \left[
 T^q(x) F^q(x)  
+ \widetilde{T}^q(x) \widetilde{F}^q(x)
 \right]
 \,,
\label{eq:factorizedamplitude}
\end{eqnarray}
where:
\begin{eqnarray}
T^q &=&
C_0^q 
+ C_1^{q}
+\log
\left(
\frac{M^2_{\gamma\gamma}}{\mu_F^2}
\right) C_{coll}^{q} 
\,,
\label{eq:coeff}
\end{eqnarray}
and the corresponding equation for the axial amplitude.
\section{The vector amplitude}
Let us first study the vector GPD contribution to the scattering amplitude.
\subsection{Lowest order analysis}
\begin{figure}[h]
    \centering
\includegraphics[width = 0.4 \textwidth]{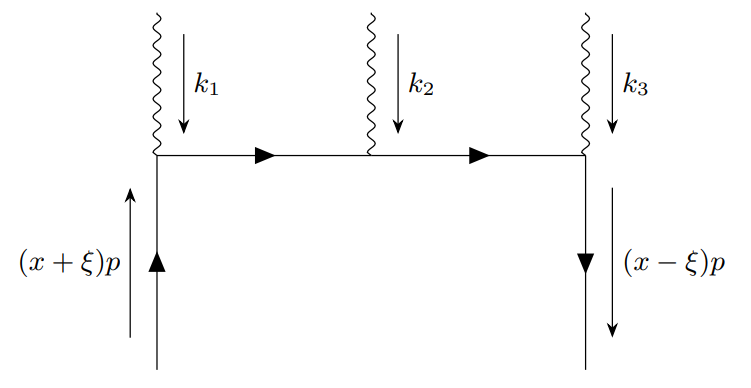}
    \caption{The general form of the LO diagram. As described in the text, to simplify computations we first consider a diagram with 3 incoming photons, and then we use the appropriate substitutions.}
    \label{fig:LO}
\end{figure}
At leading (i.e. zeroth) order in $\alpha_s$, the contribution to the vector amplitude coming from the diagram shown in Fig. \ref{fig:LO}, with a given permutation of photons,  reads~\cite{Pedrak:2017cpp}:
\begin{align}
Tr [i\mathcal{M}^0_{1,2,3}~\slashed{p}]
	= -\frac{ie_q^3}{s^{2} \beta_1 \beta_2\beta_3} \frac{1}{x+\xi +i0_1} \frac{1}{x-\xi -i0_3} 4\beta_2 \mathcal{A}_{1,2,3}\,,
	\label{eq-amp-lo-1}
\end{align}
where we denote $i0_i = i0 \cdot \mathrm{sgn}(\beta_i)$ and define the trace structure:
\begin{equation}
4\mathcal{A}_{1,2,3} := 
Tr\big( 
	\slashed{\epsilon}_3 (-\slashed{k}_3) \slashed{\epsilon}_2 \slashed{k}_1 \slashed{\epsilon}_1 \slashed{p} 
	\big)\,,
\end{equation}
 since it turns out that it appears in all diagrams. The full amplitude of the hard subprocess is obtained by summing over all permutations of photons $k_i$:
 \begin{equation}
     Tr[i\mathcal{M}^0 \slashed{p}]= \sum_{perm.}Tr [i\mathcal{M}^0_{1,2,3}\slashed{p}]\,.
 \end{equation}
 The trace structure has the important property:
 \begin{equation}
\sum_{perm.} \beta_2 \mathcal{A}_{1,2,3} =0 \,, 
 \end{equation}
which results in the vanishing of the real part of the amplitudes.  Let us observe that the interchange of indices $1\leftrightarrow 3$ in Eq. \eqref{eq-amp-lo-1} does not change it if $\mathrm{sgn}(\beta_1) = \mathrm{sgn}(\beta_3)$, and results in complex conjugation of the term $\frac{1}{x+\xi + i0_1} \frac{1}{x-\xi -i0_3}$ if $\mathrm{sgn}(\beta_1) = -\mathrm{sgn}(\beta_3)$. The only terms proportional to the imaginary part of propagators which survive, are those from diagrams in which the incoming photon is the middle one (so that $k_2 \rightarrow q$). Finally:
 \begin{equation}\label{eq-Cq0-recap}
\begin{aligned}
	&C^q_0 (x, \xi) = \frac{ie_q^3\mathcal{A} }{ M^2_{\gamma\gamma} \alpha \bar{\alpha} } \: \mathrm{Im} \Big( \frac{1}{x+\xi - i0}  \Big) + ( x \rightarrow -x)\,,
	\end{aligned}
\end{equation}
where:
\begin{equation}
\begin{aligned}	    
\mathcal{A}=	\bigg[(\alpha - \bar{\alpha} )\big( \vec{\epsilon^*}_t(\mathbf{q}_1) \vec{\epsilon^*}_t(\mathbf{q}_2) \big) \big( \vec{p}_t \vec{\epsilon}_t (\mathbf{q}) \big)  - \big(  \vec{p}_t \vec{\epsilon^*}_t(\mathbf{q}_1) \big) \big(  \vec{\epsilon}_t (\mathbf{q}) \vec{\epsilon^*}_t(\mathbf{q}_2) \big) + \big(  \vec{p}_t \vec{\epsilon^*}_t(\mathbf{q}_2) \big) \big(  \vec{\epsilon}_t(\mathbf{q}) \vec{\epsilon^*}_t(\mathbf{q}_1) \big) \bigg]  .
\end{aligned}
\end{equation}

This reproduces the result of  \cite{Pedrak:2017cpp}. 

\subsection{Self-energy insertions at $O(\alpha_s)$}

The contributions of self-energy diagrams shown in Fig. \ref{2LR} read:
\begin{figure}[h]
        \centering
\includegraphics[width = 0.8 \textwidth]{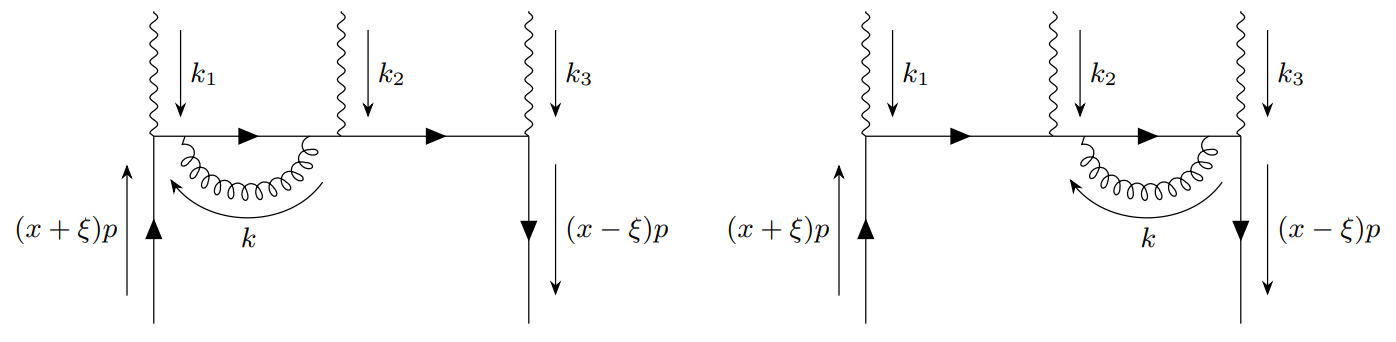}
    \caption{Self energy insertion contributions to the NLO amplitude: diagrams $2.L$ and $2.R$.}
    \label{2LR}
\end{figure}

\begin{align}
Tr[i\mathcal{M}^{2.L}_{1,2,3}~\slashed{p}]
=
Tr[i\mathcal{M}^{0}_{1,2,3}~\slashed{p}]\cdot
\frac{\alpha_S C_F}{4\pi}\Big( \frac{M^2_{\gamma\gamma} e^\gamma}{4\pi\mu^2} \Big)^{-\frac{\varepsilon}{2}}
\bigg{(} 
-\frac{2}{\varepsilon} + \log\Big(-\frac{x+\xi}{2\xi}\beta_1 -i0\Big)- 1 
\bigg{)}\,,
\label{eq:2L}
\end{align}
for the left panel diagram, and
\begin{align}
Tr[i\mathcal{M}^{2.R}_{1,2,3}~\slashed{p}]
= Tr[i\mathcal{M}^{0}_{1,2,3}~\slashed{p}]\cdot
\frac{\alpha_S C_F}{4\pi}\Big( \frac{M^2_{\gamma\gamma} e^\gamma}{4\pi\mu^2} \Big)^{-\frac{\varepsilon}{2}}
\bigg{(} 
-\frac{2}{\varepsilon} + \log\Big(\frac{x-\xi}{2\xi}\beta_3 -i0\Big)- 1 
\bigg{)}\,,
\label{eq:2R}
\end{align}
for the right panel diagram.

\subsection{Vertex corrections at $O(\alpha_s)$}

\begin{figure}[h]
    \centering
\includegraphics[width=0.8 \textwidth]{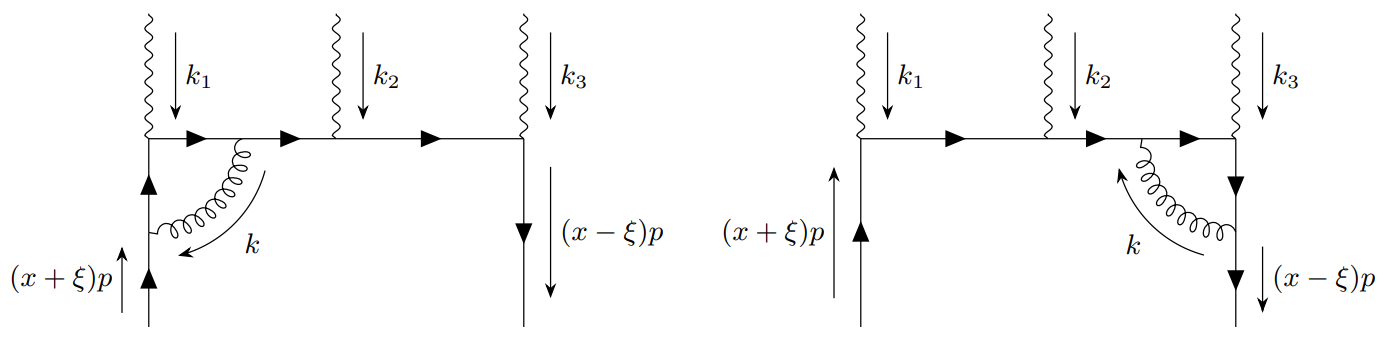}
    \caption{Vertex insertion contributions to the NLO amplitude: diagrams 3.L and 3.R.}
    \label{3LR}
\end{figure}
The left and right vertex corrections, shown in Fig. \ref{3LR}  denoted respectively as $\mathcal{M}^{3.L}$ and $\mathcal{M}^{3.R}$, read:
\begin{align}
Tr[i\mathcal{M}^{3.L}_{1,2,3}~\slashed{p}]
=Tr[i\mathcal{M}^{0}_{1,2,3}~\slashed{p}]\cdot
\frac{\alpha_S C_F}{4\pi}\Big( \frac{M^2_{\gamma\gamma} e^\gamma}{4\pi\mu^2} \Big)^{-\frac{\varepsilon}{2}}
\bigg{(} 
-\frac{2}{\varepsilon} + \log\Big(-\frac{x+\xi}{2\xi}\beta_1 -i0\Big)- 4 
\bigg{)}\,,
\label{eq:3L}
\end{align}

\begin{align}
Tr[i\mathcal{M}^{3.R}_{1,2,3}~\slashed{p}]
=Tr[i\mathcal{M}^{0}_{1,2,3}~\slashed{p}]\cdot
\frac{\alpha_S C_F}{4\pi}\Big( \frac{M^2_{\gamma\gamma} e^\gamma}{4\pi\mu^2} \Big)^{-\frac{\varepsilon}{2}}
\bigg{(} 
-\frac{2}{\varepsilon} + \log\Big(\frac{x-\xi}{2\xi}\beta_3 -i0\Big)- 4 
\bigg{)}\,,
\label{eq:3R}
\end{align}

\begin{figure}[h]
    \centering
\includegraphics[width=0.5\textwidth]{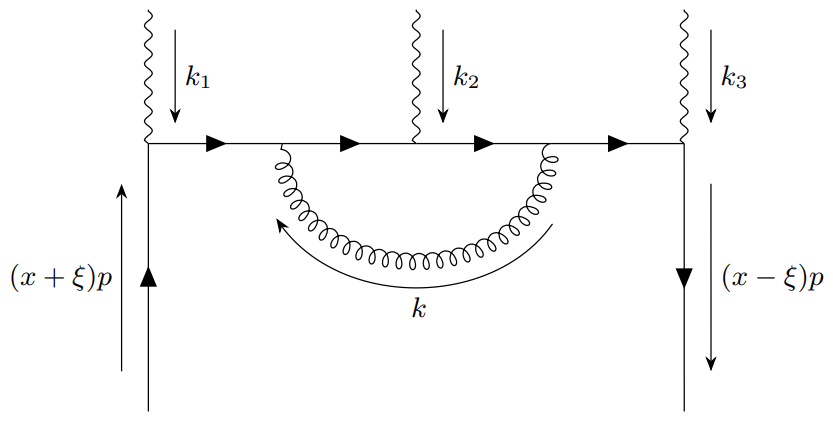}
    \caption{Vertex insertion contributions to the NLO amplitude: diagram 3.M}
    \label{3M}
\end{figure}

The middle vertex correction, shown in Fig. \ref{3M}, has a more complicated structure. Introducing $D_1= (x+\xi)\beta_1$, $D_3= -(x-\xi)\beta_3$, we can write it as:
\begin{align}
Tr[i\mathcal{M}^{3.M}_{1,2,3}~\slashed{p}]
=&
Tr[i\mathcal{M}^{0}_{1,2,3}~\slashed{p}]\cdot
\frac{\alpha_S C_F}{4\pi}\Bigg\{\Big( \frac{M^2_{\gamma\gamma} e^\gamma}{4\pi\mu^2} \Big)^{-\frac{\varepsilon}{2}}
\bigg{(} 
+\frac{2}{\varepsilon} +1- \frac{D_1\log\Big(\frac{-D_1 -i0}{2\xi}\Big) -  D_3\log\Big(\frac{-D_3 -i0}{2\xi}\Big)}{D_1-D_3}
\bigg{)} 
\nonumber\\
&
\phantom{Tr[i\mathcal{M}^{0}_{1,2,3}~\slashed{p}]\cdot
\frac{\alpha_S C_F}{4\pi}\Bigg\{}
+\bigg(  \frac{\mathcal{B}_{1,2,3}}{\mathcal{A}_{1,2,3}}\bigg)
\frac{D_1}{D_1-D_3}
\left(
1+\frac{-2D_1+D_3}{D_1-D_3}\log\frac{D_1+i0}{D_3+i0}
\right)
\nonumber\\
&
\phantom{Tr[i\mathcal{M}^{0}_{1,2,3}~\slashed{p}]\cdot
\frac{\alpha_S C_F}{4\pi}\Bigg\{}
+\bigg(\frac{\mathcal{C}_{1,2,3}}{\mathcal{A}_{1,2,3}}\bigg)
\frac{D_3}{D_1-D_3}
\left(
1+\frac{D_1-2D_3}{D_1-D_3}\log\frac{D_1+i0}{D_3+i0}
\right)
\Bigg\}\,,
\label{eq:3M}
\end{align}
where:
\begin{align}
    \mathcal{B}_{1,2,3} = Tr\Big{\{} \slashed{\epsilon}_3 \slashed{k}_3 \slashed{k}_1 \slashed{\epsilon}_2\slashed{p} \slashed{\epsilon}_1 \Big{\}}& \nonumber\,,~~~~
    \mathcal{C}_{1,2,3} =Tr\Big{\{} \slashed{\epsilon}_3 \slashed{p}\slashed{\epsilon}_2  \slashed{k}_1 \slashed{k}_3\slashed{\epsilon}_1 \Big{\}}. &
\end{align}

\subsection{The box contributions }
\begin{figure}[h]
    \centering
\includegraphics[width=0.8 \textwidth]{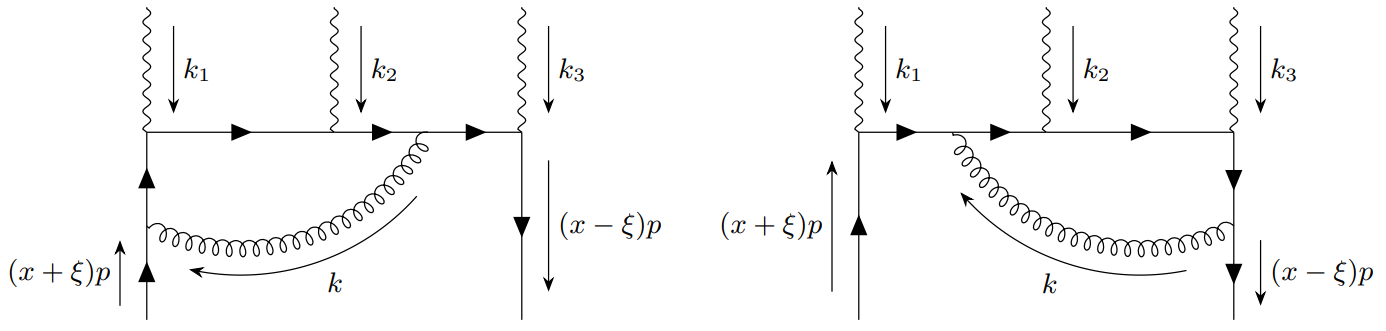}
    \caption{Box contributions to the NLO amplitude: diagrams 4.L and 4.R.}
    \label{4LR}
\end{figure}
There are two box contributions shown on Fig. \ref{4LR}. The divergent part of the amplitude $4.L$ reads:
\begin{equation}
\begin{aligned}
    Tr\big[i\mathcal{M}^{4.L}_{1,2,3}~ \slashed{p} \big]_{\mathrm{div.}}&= 
-\frac{4}{\varepsilon} \frac{\alpha_S C_F}{4\pi} Tr\big[i\mathcal{M}^{0}_{1,2,3}~ \slashed{p} \big] \log \Big{(} -\frac{x-\xi}{2\xi} + i0_3 \Big{)}\,,
\end{aligned}
\label{eq:4L}
\end{equation}
while the divergent part of the second box diagram reads:
\begin{equation}
\begin{aligned}
    Tr\big[i\mathcal{M}^{4.R}_{1,2,3}~\slashed{p}\big]_{\mathrm{div}}&= 
-\frac{4}{\varepsilon} \frac{\alpha_S C_F}{4\pi} Tr\big[i\mathcal{M}^{0}_{1,2,3}~ \slashed{p} \big] \log \Big{(} \frac{x+\xi}{2\xi} + i0_1 \Big{)}.
\end{aligned}
\label{eq:4R}
\end{equation}
The finite contribution resulting from these box diagrams is lengthy and will be  included in formulas presented in Sec. \ref{sec:finite}.

\subsection{The pentagonal contribution}
\begin{figure}[h]
\includegraphics[width=0.5 \textwidth]{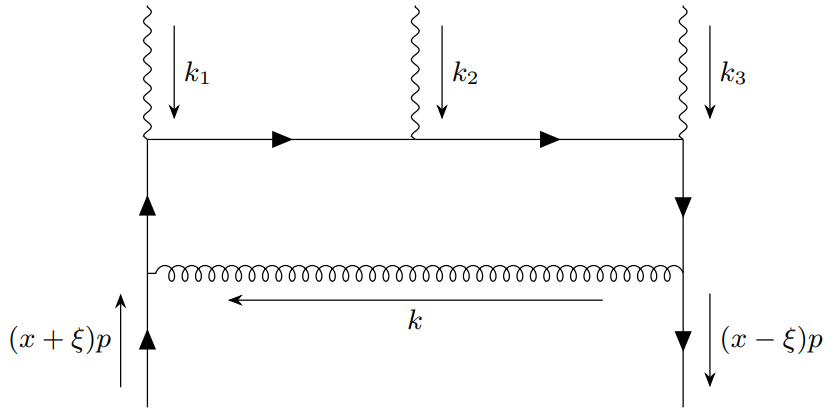}
    \caption{The 5-point diagram contributing to the NLO amplitude.}
    \label{5}
\end{figure}
The pentagonal contribution to the amplitude, resulting from the diagram presented on Fig. \ref{5}, reads:
\begin{equation}
Tr\big[ i\mathcal{M}^5_{1,2,3}~ \slashed{p} \big] = C_F e_q^3 g^2 \mu^{4-d} \int \frac{d^d k}{(2\pi)^d} \frac{Tr\Big{\{} \gamma_\mu \slashed{p} \gamma^\mu (\slashed{\underline{p}} + \slashed{k})\slashed{\epsilon}_3(\slashed{\bar{p}} + \slashed{k}_1 + \slashed{k}_2 + \slashed{k})\slashed{\epsilon}_2 (\slashed{\bar{p}} + \slashed{k}_1 + \slashed{k}) \slashed{\epsilon}_1 (\slashed{\bar{p}} + \slashed{k}) \Big{\}}}{(k^2+i0)(D_+ +i0)(D_{-} +i0)(D_1+i0) (D_{12}+i0)},
\end{equation}
where $\bar{p} = (x+\xi)p$, $\underline{p} = (x-\xi)p$ and:
$$
D_{\pm} = \big{(} (x\pm \xi)p + k  \big{)}^2, \quad D_{1} = \big{(} (x + \xi)p + k_1 + k  \big{)}^2, \quad D_{12} = \big{(} (x + \xi)p + k_1 + k_2 + k  \big{)}^2.
$$
Observing that
$$
2(pk) = \frac{1}{2\xi} (D_+ - D_{-} ), \quad k^2 = \frac{1}{2\xi} \big{(} (x+\xi)D_{-} - (x-\xi)D_+ \big{)}\,,
$$
and using the following gamma-identities inside the trace:
$
\gamma^\mu \slashed{p} \gamma_\mu = (-2 + \varepsilon) \slashed{p}, \quad \slashed{k} \slashed{p} \slashed{k} = 2(pk) \slashed{k} - k^2 \slashed{p}
$, allows us to obtain the following identity:
\begin{equation}
2(pk) \slashed{k} - k^2 \slashed{p} = \frac{D_+}{2\xi} (\slashed{k}+\slashed{\underline{p}}) - \frac{D_{-}}{2\xi} (\slashed{k} + \slashed{\bar{p}}),
\end{equation}
and to write the  amplitude  as the sum of two 4-point integrals:
\begin{equation}
Tr\big[ i\mathcal{M}^5_{1,2,3}~\slashed{p}\big] = Tr\big[ i\mathcal{M}^{5.L}_{1,2,3}~\slashed{p} \big] + Tr\big[ i\mathcal{M}^{5.R}_{1,2,3}~\slashed{p} \big] ,
\end{equation}
where
\begin{align}
&Tr\big[ i\mathcal{M}^{5.L}_{1,2,3}~ \slashed{p}\big] := - \frac{-2+\varepsilon}{2\xi}C_F e_q^3 g^2 \mu^{4-d} \int \frac{d^d k}{(2\pi)^d} \frac{Tr\Big{\{} (\slashed{\bar{p}} + \slashed{k})\slashed{\epsilon}_3(\slashed{\bar{p}} + \slashed{k}_1 + \slashed{k}_2 + \slashed{k})\slashed{\epsilon}_2 (\slashed{\bar{p}} + \slashed{k}_1 + \slashed{k}) \slashed{\epsilon}_1 \Big{\}}}{(k^2+i0)(D_{+} +i0)(D_1+i0) (D_{12}+i0)}, \label{eq-M-5-split}
\\
&Tr\big[ i\mathcal{M}^{5.R}_{1,2,3}~\slashed{p} \big] := \frac{-2+\varepsilon}{2\xi}C_F e_q^3 g^2 \mu^{4-d} \int \frac{d^d k}{(2\pi)^d} \frac{Tr\Big{\{} (\slashed{\underline{p}} + \slashed{k})\slashed{\epsilon}_3(\slashed{\bar{p}} + \slashed{k}_1 + \slashed{k}_2 + \slashed{k})\slashed{\epsilon}_2 (\slashed{\bar{p}} + \slashed{k}_1 + \slashed{k}) \slashed{\epsilon}_1 \Big{\}}}{(k^2+i0)(D_{-} +i0)(D_1+i0) (D_{12}+i0)}.
\end{align}
The names $5.L$ and $5.R$ come from the structure of denominators in integrands, which are the same as in amplitudes $4.L$ and $4.R$. The divergent part of the amplitude reads:

\begin{equation}
\begin{aligned}
	&Tr\big[ i\mathcal{M}^5_{1,2,3}~\slashed{p}\big]_{\mathrm{div.}} =- \frac{\alpha_S C_F}{4\pi} Tr\big[ i\mathcal{M}^0_{1,2,3}~\slashed{p}\big]
\frac{2}{\varepsilon} 
\bigg(  \frac{x-\xi}{\xi}\log \Big{(}- \frac{x -\xi}{2\xi}+ i0_3\big{)} \Big{)} 
- \frac{x+\xi}{\xi}\log \Big{(} \frac{x+\xi}{2\xi} + i0_1 \Big{)} \bigg)\,.
\end{aligned}
\label{eq:5}
\end{equation}
The finite contribution resulting from this pentagonal diagram will be included in formulas presented in Sect. \ref{sec:finite}.

\subsection{Total divergent part}
Summing the various contributions written above in Eqs. (\ref{eq:2L}, \ref{eq:2R}, \ref{eq:3L}, \ref{eq:3R}, \ref{eq:3M}, \ref{eq:4L}, \ref{eq:4R}) and (\ref{eq:5})
we can write the divergent part of the amplitude corresponding to a given permutation of photons entering the hard part as:
\begin{equation}
\begin{aligned}
	&Tr\big[ i\mathcal{M}_{1,2,3}~\slashed{p}\big]_{\mathrm{div.}} = -\frac{2}{\varepsilon} \cdot \frac{\alpha_S C_F}{4\pi} \cdot Tr\big[ i\mathcal{M}^0_{1,2,3}~\slashed{p}\big]
\bigg( 3
+\frac{x+\xi}{\xi}\log \Big{(} 
        -\frac{x-\xi}{2\xi} + i0_3 \Big{)} 
- \frac{x-\xi}{\xi}\log \Big{(} \frac{x+\xi}{2\xi} + i0_1\big{)} \Big{)} \bigg) \,,
\end{aligned}
\end{equation}
which after the summation over permutations gives (see the definition \eqref{eq-def-Ccoll}):
\begin{equation}\label{eq-Cqcoll-fin}
\begin{aligned}
&C^q_{coll} = \frac{i e_q^3 \mathcal{A}}{ M^2_{\gamma\gamma} \alpha \bar{\alpha} } \frac{\alpha_S C_F}{4\pi} 
&  \mathrm{Im} \bigg[ \frac{1}{x+\xi-i0} \bigg( -3- 2
\log \Big{(} \frac{x+\xi}{2\xi} - i0 \Big{)}  \bigg) \bigg] + (x \rightarrow -x)\,.
\end{aligned}
\end{equation}
The resulting expressions for $C_0^q$ and $C_{coll}^q$ are, up to a common constant factor, equal to the imaginary part of the corresponding expressions in DVCS (see Eq. (13) in \cite{Moutarde:2013qs}). The non-singlet and singlet kernels $K_{qq}$ are at the $\alpha_S$ order equal to each other \cite{Diehl:2003ny} and real, hence the condition for the cancellation of divergences expressed by Eq. (\ref{eq:div_cancel}) in the case of the considered process is just the imaginary part of the corresponding expression for DVCS. 

\subsection{The finite part of $C_1$}
\label{sec:finite}

The finite parts of the contributions of self energy and vertex diagrams to $C_1$ are explicitly written in Eqs. \eqref{eq:2L}-\eqref{eq:3M}. The finite parts emerging from the box and pentagon diagrams are more cumbersome and  it is convenient to write them in terms of functions defined by the following equations:
\begin{align}
(-1)^n s^{-n} \mathcal{F}_{nab}\big(x, \xi, \{ \beta_i \} \big) &:=
\int_0^1 dr dz \: r^{a}z^{b}\Big{(} -2\big{(} z\bar{p}k_1 + zr \bar{p}k_2 + r k_1 k_2 \big{)} - i0  \Big{)}^{-n}\,,
\nonumber \\
&= (-1)^n s^{-n} \int_0^1 dr dz \: r^{a}z^{b}\Big{(} \big{(} z(x+\xi)\beta_1 + zr (x+\xi)\beta_2 + r \kappa_3 \big{)} + i0  \Big{)}^{-n}\,, 
\end{align}
\begin{equation}
	\mathcal{G}(x, \xi, \{ \beta_i \}) := \int_0^1 dr dz\: z\Big{(} \big{(} z(x+\xi)\beta_1 + zr(x+\xi)\beta_2 + r\kappa_3 + i0  \Big{)}^{-2}\ln \big(- z(x+\xi)\beta_1 - zr(x+\xi)\beta_2 - r\kappa_3 - i0 \big)\,.
\end{equation}
It is worth stressing that the integrals in the above equations are finite. Using above definitions,  
and omitting for brevity the argument of functions $\mathcal{F}_{nab}$ and $\mathcal{G}$,
we can write the finite part of the box diagram to which this permutation contributes as:

\begin{equation}\label{eq-finite-4L}
\begin{aligned}
&Tr\big[ i\mathcal{M}^{4.L}_{1,2,3} ~\slashed{p} \big]_{\mathrm{fin.}} = -i \frac{\alpha_SC_F}{4\pi} e_q^3  \frac{1}{2\underline{p}k_3 - i0} s^{-2} \bigg\{ \\
&\phantom{Aa}\mathcal{F}_{210}~ \Big( Tr\big( \slashed{k}_3 \slashed{\epsilon}_3 \slashed{p} \slashed{k}_1 \slashed{\epsilon}_2 \slashed{k}_2 \slashed{\epsilon}_1 \slashed{k}_1 \big) - Tr\big( \slashed{k}_3 \slashed{\epsilon}_3 \slashed{p} \slashed{k}_1 \slashed{\epsilon}_2 \slashed{k}_1 \slashed{\epsilon}_1 \slashed{k}_2 \big) -2Tr\big( \slashed{k}_3 \slashed{\epsilon}_3 \slashed{p} \slashed{k}_2 \slashed{\epsilon}_2 \slashed{k}_1 \slashed{\epsilon}_1 \slashed{k}_2 \big) \Big)\\
&+\mathcal{F}_{201}~ \Big( - Tr\big( \slashed{k}_3 \slashed{\epsilon}_3 \slashed{p} \slashed{k}_1 \slashed{\epsilon}_2 \bar{\slashed{p}} \slashed{\epsilon}_1 \slashed{k}_1 \big) -2Tr\big( \slashed{k}_3 \slashed{\epsilon}_3 \slashed{p} \slashed{k}_2 \slashed{\epsilon}_2 \bar{\slashed{p}} \slashed{\epsilon}_1 \slashed{k}_1 \big) 
- 4 Tr\big( \slashed{k}_3 \slashed{\epsilon}_3 \slashed{p} \slashed{k}_2 \slashed{\epsilon}_2 \slashed{k}_1 \slashed{\epsilon}_1 \bar{\slashed{p}} \big) - 5 Tr\big( \slashed{k}_3 \slashed{\epsilon}_3 \slashed{p} \slashed{k}_1 \slashed{\epsilon}_2 \slashed{k}_1 \slashed{\epsilon}_1 \bar{\slashed{p}} \big) \Big)\\
&+\mathcal{F}_{211}~\Big( -Tr\big( \slashed{k}_3 \slashed{\epsilon}_3 \slashed{p} \slashed{k}_1 \slashed{\epsilon}_2 \bar{\slashed{p}} \slashed{\epsilon}_1 \slashed{k}_2 \big) - Tr\big( \slashed{k}_3 \slashed{\epsilon}_3 \slashed{p} \slashed{k}_1\slashed{\epsilon}_2 \slashed{k}_2 \slashed{\epsilon}_1 \bar{\slashed{p}} \big) 
+ Tr\big( \slashed{k}_3 \slashed{\epsilon}_3 \slashed{p} \slashed{k}_2 \slashed{\epsilon}_2 \bar{\slashed{p}} \slashed{\epsilon}_1 \slashed{k}_1 \big) 
\\
&\phantom{AAAAAA}-2 Tr\big( \slashed{k}_3 \slashed{\epsilon}_3 \slashed{p} \slashed{k}_2 \slashed{\epsilon}_2 \bar{\slashed{p}} \slashed{\epsilon}_1 \slashed{k}_2 \big) -  Tr\big( \slashed{k}_3 \slashed{\epsilon}_3 \slashed{p} \slashed{k}_2 \slashed{\epsilon}_2 \slashed{k}_1 \slashed{\epsilon}_1 \bar{\slashed{p}} \big) \Big) \\
&+\mathcal{F}_{220}~\Big( Tr\big( \slashed{k}_3 \slashed{\epsilon}_3 \slashed{p} \slashed{k}_1 \slashed{\epsilon}_2 \slashed{k}_2 \slashed{\epsilon}_1 \slashed{k}_2 \big) + Tr\big( \slashed{k}_3 \slashed{\epsilon}_3 \slashed{p} \slashed{k}_2 \slashed{\epsilon}_2 \slashed{k}_1 \slashed{\epsilon}_1 \slashed{k}_2 \big) \Big) 
+\mathcal{F}_{221}~Tr\big( \slashed{k}_3 \slashed{\epsilon}_3 \slashed{p} \slashed{k}_2 \slashed{\epsilon}_2 \bar{\slashed{p}} \slashed{\epsilon}_1 \slashed{k}_2 \big)\\
& +s \mathcal{F}_{100}~ \Big( Tr\big(\slashed{k}_3 \slashed{\epsilon}_3 \slashed{p} \slashed{\epsilon}_2 \slashed{\epsilon}_1 \slashed{k}_1 \big) -  Tr\big(\slashed{p}  \slashed{\epsilon}_3 \slashed{k}_3 \slashed{\epsilon}_2 \slashed{k}_1 \slashed{\epsilon}_1  \big) - Tr\big(\slashed{k}_3 \slashed{\epsilon}_3 \slashed{p} \slashed{k}_1 \slashed{\epsilon}_2 \slashed{\epsilon}_1 \big) -2Tr\big(\slashed{k}_3 \slashed{\epsilon}_3 \slashed{p} \slashed{k}_2 \slashed{\epsilon}_2 \slashed{\epsilon}_1 \big) \Big)  \\
&+s \mathcal{F}_{110}~\Big( Tr\big(\slashed{k}_3 \slashed{\epsilon}_3 \slashed{p} \slashed{\epsilon}_2 \slashed{\epsilon}_1 \slashed{k}_2 \big) + Tr\big(\slashed{p}  \slashed{\epsilon}_3 \slashed{k}_3 \slashed{\epsilon}_2 \slashed{k}_2 \slashed{\epsilon}_1  \big)+ Tr\big(\slashed{k}_3 \slashed{\epsilon}_3 \slashed{p} \slashed{k}_2 \slashed{\epsilon}_2 \slashed{\epsilon}_1 \big) \Big)
+ 2 \mathcal{G} Tr\big( \slashed{k}_3 \slashed{\epsilon}_3 \slashed{p} \slashed{k}_3 \slashed{\epsilon}_2 \slashed{k}_1 \slashed{\epsilon}_1 \bar{\slashed{p}} \big) \Big) \bigg\}\\
& - \frac{\alpha_SC_F}{2\pi} \cdot Tr\big[ i\mathcal{M}^0_{1,2,3}~\slashed{p}\big] \log(2\xi) \log \Big{(} \frac{1}{2\xi}\big{(}\xi - x + i0\cdot \mathrm{sgn}(\beta_3)\big{)} \Big{)}
\end{aligned}
\end{equation}
The finite part of the pentagonal diagram reads:
\begin{equation}\label{eq-finite-5L}
\begin{aligned}
&Tr\big[ i\mathcal{M}^{5.L}_{1,2,3} ~\slashed{p} \big]_{\mathrm{fin.}} =-i\frac{\alpha_S C_F}{4\pi} e_q^3  \frac{1}{\xi}s^{-2}\frac{1}{2} \bigg\{ \\
&\phantom{Aa}\mathcal{F}_{201}~\Big( Tr\big( \slashed{k}_1 \slashed{\epsilon}_3 \bar{\slashed{p}} \slashed{\epsilon}_2 \slashed{k}_1 \slashed{\epsilon}_1  \big) + Tr\big( \slashed{k}_1 \slashed{\epsilon}_3 \slashed{k}_1 \slashed{\epsilon}_2  \bar{\slashed{p}} \slashed{\epsilon}_1  \big) 
	+ 2 Tr\big( \slashed{k}_1 \slashed{\epsilon}_3 \slashed{k}_2 \slashed{\epsilon}_2  \bar{\slashed{p}} \slashed{\epsilon}_1  \big)  + \\
& \phantom{AAAAAAAA}+ 5 Tr\big( \bar{\slashed{p}} \slashed{\epsilon}_3 \slashed{k}_1 \slashed{\epsilon}_2 \slashed{k}_1 \slashed{\epsilon}_1 \big) + 4 Tr\big( \bar{\slashed{p}} \slashed{\epsilon}_3 \slashed{k}_2 \slashed{\epsilon}_2 \slashed{k}_1 \slashed{\epsilon}_1 \big) \Big)  \\
&+\mathcal{F}_{210}~\Big( -Tr\big( \slashed{k}_1 \slashed{\epsilon}_3 \slashed{k}_1 \slashed{\epsilon}_2  \slashed{k}_2 \slashed{\epsilon}_1  \big) +Tr\big( \slashed{k}_2 \slashed{\epsilon}_3 \slashed{k}_1 \slashed{\epsilon}_2 \slashed{k}_1 \slashed{\epsilon}_1  \big) 
	+2Tr\big( \slashed{k}_2 \slashed{\epsilon}_3 \slashed{k}_2 \slashed{\epsilon}_2 \slashed{k}_1 \slashed{\epsilon}_1  \big) \Big)  \\
&+ \mathcal{F}_{211}~ \Big( Tr\big( \bar{\slashed{p}} \slashed{\epsilon}_3 \slashed{k}_1 \slashed{\epsilon}_2 \slashed{k}_2 \slashed{\epsilon}_1  \big) + Tr\big( \bar{\slashed{p}} \slashed{\epsilon}_3 \slashed{k}_2 \slashed{\epsilon}_2 \slashed{k}_1 \slashed{\epsilon}_1  \big) - Tr\big( \slashed{k}_1 \slashed{\epsilon}_3 \bar{\slashed{p}} \slashed{\epsilon}_2 \slashed{k}_2 \slashed{\epsilon}_1  \big)   \\
&\phantom{AAAAAAAA} - Tr\big( \slashed{k}_1 \slashed{\epsilon}_3 \slashed{k}_2 \slashed{\epsilon}_2  \bar{\slashed{p}} \slashed{\epsilon}_1  \big) + Tr\big( \slashed{k}_2 \slashed{\epsilon}_3 \bar{\slashed{p}} \slashed{\epsilon}_2 \slashed{k}_1 \slashed{\epsilon}_1  \big) +
 Tr\big( \slashed{k}_2 \slashed{\epsilon}_3 \slashed{k}_1 \slashed{\epsilon}_2 \bar{\slashed{p}} \slashed{\epsilon}_1  \big) + 2 Tr\big( \slashed{k}_2 \slashed{\epsilon}_3 \slashed{k}_2 \slashed{\epsilon}_2 \bar{\slashed{p}} \slashed{\epsilon}_1  \big) \Big)  \\
&+  \mathcal{F}_{220}~ \Big( -Tr\big( \slashed{k}_2 \slashed{\epsilon}_3 \slashed{k}_1 \slashed{\epsilon}_2 \slashed{k}_2 \slashed{\epsilon}_1  \big) - Tr\big( \slashed{k}_2 \slashed{\epsilon}_3 \slashed{k}_2 \slashed{\epsilon}_2 \slashed{k}_1 \slashed{\epsilon}_1  \big) \Big) 
-  \mathcal{F}_{221}~ \Big( Tr\big( \slashed{k}_2 \slashed{\epsilon}_3 \bar{\slashed{p}} \slashed{\epsilon}_2 \slashed{k}_2 \slashed{\epsilon}_1  \big) + Tr\big( \slashed{k}_2 \slashed{\epsilon}_3 \slashed{k}_2 \slashed{\epsilon}_2 \bar{\slashed{p}} \slashed{\epsilon}_1  \big) \Big) \\
&+\mathcal{F}_{100}~\Big( Tr\big( \slashed{\epsilon}_3 \slashed{\epsilon}_2 \slashed{k}_1 \slashed{\epsilon}_1 \big) + Tr\big( \slashed{\epsilon}_3 \slashed{k}_1 \slashed{\epsilon}_2  \slashed{\epsilon}_1 \big) - Tr\big( \slashed{\epsilon}_3 \slashed{\epsilon}_2  \slashed{\epsilon}_1 \slashed{k}_1 \Big) + 2 Tr\big( \slashed{\epsilon}_3 \slashed{k}_2 \slashed{\epsilon}_2  \slashed{\epsilon}_1 \big) \Big) \\
&-\mathcal{F}_{110}\Big(  Tr\big( \slashed{\epsilon}_3  \slashed{\epsilon}_2 \slashed{k}_2 \slashed{\epsilon}_1 \big) + Tr\big( \slashed{\epsilon}_3 \slashed{k}_2 \slashed{\epsilon}_2  \slashed{\epsilon}_1 \big) + Tr\big( \slashed{\epsilon}_3  \slashed{\epsilon}_2  \slashed{\epsilon}_1 \slashed{k}_2\big) \Big) 
+2 \mathcal{G} Tr\big( \bar{\slashed{p}} \slashed{\epsilon}_3 \slashed{k}_3 \slashed{\epsilon}_2 \slashed{k}_1 \slashed{\epsilon}_1 \big)  \\
&+2 \mathcal{A} \frac{s^2 (x-\xi)}{(\underline{p}k_3-i0)(\bar{p}k_1+i0)} \log \Big{(} \frac{1}{2\xi}\big{(}\xi - x + i0\cdot \mathrm{sgn}(pk_3)\big{)} \Big{)}\bigg\}\\
&- \frac{\alpha_S}{4\pi}C_F \cdot Tr\big[ i\mathcal{M}^0_{1,2,3}~\slashed{p}\big] \log(2\xi) \frac{x-\xi}{\xi}\log \Big{(} \frac{1}{2\xi}\big{(}\xi - x + i0\cdot \mathrm{sgn}(\beta_3)\big{)} \Big{)}.
\end{aligned}
\end{equation}
Terms in the last lines of Eqs. \eqref{eq-finite-4L} and \eqref{eq-finite-5L} are not present in the work \cite{Grocholski:2021gta} - they result from using different definitions of the hard scale ($M_{\gamma\gamma}^2$ here versus $s$ in \cite{Grocholski:2021gta}). The finite parts of diagrams $4.R$  and $5.R$ are obtained from \eqref{eq-finite-4L} and \eqref{eq-finite-5L} respectively by substituting $k_1 \rightarrow -k_3$, $k_2 \rightarrow -k_2$, $k_3 \rightarrow -k_1$, $\xi \rightarrow -\xi$, and $\epsilon_1 \leftrightarrow \epsilon_3$.

 At NLO, the summation over photons' permutations does not yield as many simplifications as it did at the leading order, so that the calculation of the finite part will be performed numerically with the help of the PARTONS framework \cite{Berthou:2015oaw}.
\section{The axial amplitude}
We shall not detail the corresponding analysis for the axial amplitude, which follows the same lines as the vector part, but only quote the main results. The axial LO amplitude corresponding to a given permutation of photons reads~\cite{Pedrak:2017cpp}:
\begin{equation}
    Tr\Big[ i\mathcal{M}^0_{1,2,3} \gamma^5 \slashed{p} \Big] = 
    \frac{ie_q^3}{s^2\beta_1\beta_3} \frac{1}{(x+\xi)+ i0_1}\frac{1}{(x-\xi) - i0_3} 
    Tr\big( \slashed{\epsilon}_3 \slashed{k}_3 \slashed{\epsilon}_2 \slashed{k}_1 \slashed{\epsilon}_1 \gamma^5 \slashed{p} \big).
\end{equation}

After summation over all permutations of photons one gets:
\begin{equation}
\begin{aligned}
\widetilde{C}_0  = -\frac{ie_q^3\widetilde{\mathcal{A}} }{ M^2_{\gamma\gamma} \alpha \bar{\alpha} } \: \mathrm{Im} \Big( \frac{1}{x+\xi - i0}  \Big) - ( x \rightarrow -x)
\end{aligned}
\end{equation}
where:
\begin{align}
\widetilde{\mathcal{A}} = \frac{1}{2s}\Big( \bar{\alpha} Tr\big( \slashed{\epsilon}(q) \slashed{q} \slashed{\epsilon}^*(q_2) \slashed{q}_1 \slashed{\epsilon}^*(q_1) \gamma^5 \slashed{p} \big) + {\alpha} Tr\big( \slashed{\epsilon}(q) \slashed{q} \slashed{\epsilon}^*(q_1) \slashed{q}_2 \slashed{\epsilon}^*(q_2) \gamma^5 \slashed{p} \big) \Big)
\end{align}

At the first order in $\alpha_s$, an analysis very similar to the one discussed above for the vector part, yields a  divergent  part equal to:
\begin{align}
\widetilde{C}_{coll} = \frac{ie_q^3 \widetilde{\mathcal{A}}}{ M^2_{\gamma\gamma} \alpha \bar{\alpha} } \frac{\alpha_S C_F}{4\pi} 
&  \mathrm{Im} \bigg[ \frac{1}{x+\xi-i0} \bigg( -3- 2
\log \Big{(} \frac{x+\xi}{2\xi} - i0 \Big{)}  \bigg) \bigg] - (x \rightarrow -x)
\end{align}
which allows to check  factorization by explicit computation, the coefficient function divergences being exactly cancelled by those related to the evolution equations of the axial GPDs. 

Since the smallness of the axial contribution has been demonstrated in \cite{Pedrak:2017cpp} for the LO computation,  we do not write down here the lengthy expressions for the finite part of the NLO axial amplitude. They have an analogous structure as those in the vector part, including the presence of the same functions $\mathcal{F}_{nab}$ and $\mathcal{G}$, the only difference being the traces.

\section{Conclusion}
Our results on the NLO analysis of the scattering amplitude of process \eqref{process} is a first step toward a complete proof of QCD collinear factorization for the amplitude of such processes where the coefficient function describes a $2\to 3$ hard scattering.   We shall soon continue this study by examining the phenomenological consequences of our NLO results, both for reactions at electron machines at medium (JLab) or high energy \cite{AbdulKhalek:2021gbh,Anderle:2021wcy} and in ultraperipheral reactions at LHC~\cite{Pire:2008ea}. 
\paragraph*{Acknowledgements.}
\noindent
The work of O.G. is financed by the budget for science in 2020-2021, as a research project under the "Diamond Grant" program.
 The works of L.S. and J.W. are  supported respectively by the grants 2019/33/B/ST2/02588 and 2017/26/M/ST2/01074 of the National Science Center in Poland.  
This work is also partly supported by the Polish-French collaboration agreements Polonium, by the Polish National Agency for Academic Exchange and COPIN-IN2P3 and by the European Union’s Horizon 2020 research and innovation programme under grant agreement No 824093.
\bibliography{NLO_short}
\end{document}